\documentclass[12pt]{article}

\usepackage{amssymb}
\usepackage{amsmath}
\usepackage[amsmath,thmmarks]{ntheorem}

\usepackage{mathpartir}
\usepackage[latin1]{inputenc}
\usepackage{verbatim}
\usepackage{color} 

\allowdisplaybreaks

\newcommand{\impl}{\Longrightarrow}

\newcommand{\seq}{\Rightarrow}
\newcommand{\lam}[2]{\lambda#1{.}#2}
\newcommand{\siml}{\sim_\lambda}
\newcommand*{\mci}{\mathcal{I}}
\newcommand{\Subst}[3]{#1{\hskip1pt}^{#2}_{#3}}
\newcommand{\True}{1}

\newcommand{\FV}{\mathcal{V}}

\newcommand{\bool}{\ensuremath{B}}
\newcommand{\dd}[2]{\ensuremath{#1_1,\dots,#1_{#2}}}  

\newcommand{\Cond}[3]{\mathsf{if}\;#1\;\mathsf{then}\;#2\;\mathsf{else}\;#3}

\newcommand{\Quote}{{\downarrow}}
\newcommand{\valu}{c}
\newcommand{\valv}{a}
\newcommand{\valw}{b}
\newcommand{\deq}{\doteq}
\newcommand*{\mcb}{\mathcal{B}}
\newcommand{\fto}[2]{[#1\to #2]}

\theoremstyle{plain}
\newtheorem{thm}{Theorem}
\newtheorem{lem}[thm]{Lemma}
\newtheorem{prop}[thm]{Proposition}

\theoremstyle{nonumberplain}
\theoremheaderfont{\itshape}
\theorembodyfont{\normalfont}
\theoremsymbol{\ensuremath{_\square}}
\newtheorem{pf}{Proof}

\newcommand{\N}[1]{{\color{blue}#1}}
\renewcommand{\emph}[1]{\textsl{\N{#1}}}

\begin{document}

\title{A Minimal Propositional Type Theory}

\author{Mark Kaminski and Gert Smolka\\
\small Saarland University, Saarbrücken, Germany}

\maketitle{\thispagestyle{empty}}

\begin{abstract}
  Propositional type theory, first studied by Henkin,
  is the restriction of simple type theory to a single
  base type that is interpreted as the set of the two
  truth values.  We show that two constants (falsity
  and implication) suffice for denotational and
  deductive completeness.  Denotational completeness
  means that every value of the full set-theoretic type
  hierarchy can be described by a closed term.
  Deductive completeness is shown for a sequent-based
  proof system that extends a propositional natural
  deduction system with lambda conversion and Boolean
  replacement.
\end{abstract}

\section{Introduction}

Propositional type theory, first studied by
Henkin~\cite{Henkin63}, is the restriction of simple
type theory~\cite{Church40,Andrews} to a single base
type that is interpreted as the set of the two truth
values.  Functional types are interpreted as the full
set-theoretic function spaces.  As logical constants
Henkin takes all identity predicates, which
suffices to express
the propositional connectives and the quantifiers.  His
deductive system is a Hilbert system whose inference
rules are $\beta$ and replacement of equals with
equals.  Henkin first shows that his language is
denotationally complete, meaning that every value of
the full set-theoretic type hierarchy can be denoted by
a closed term. He then exploits denotational
completeness to show deductive completeness.  Deciding
validity of formulas in Henkin's language requires
nonelementary time~\cite{Vorobyov04}.

It turns out that only two constants, falsity and
implication, suffice for denotational
completeness~\cite{KaminskiSmolka08pttlambda}.  This
result raises the question for a likewise minimal proof
system.  We answer this question in this paper.  As a
basis we take a sequent-based natural deduction system
for propositional logic with falsity and implication.
We show that with two additional rules one obtains a
complete deduction system.  The first rule incorporates
lambda conversion ($\alpha$, $\beta$,~$\eta$).  The
second rule provides for Boolean replacement of equals
with equals where variable capture is admissible in
some cases.

In a previous paper~\cite{KaminskiSmolka08pttlambda},
we give a complete proof system for propositional type
theory with falsity and implication.  This system is an
instance of an equational proof system for the pure
lambda calculus.  As such it is far from minimal since
it provides for equations and replacement at all types.
In contrast, the system of the present paper has
equations and replacement only for the type of truth
values.

The paper is organized as follows.  We start by
defining the syntax and semantics of our language.
Then we show denotational completeness.  Finally we
establish a sequent-based proof system and show its
completeness.

\section{Terms, Formulas, and Validity}

We assume familiarity with the simply typed lambda
calculus (see, e.g.,~\cite{HindleySeldin}).
\emph{Types} ($\sigma$, $\tau$, $\rho$) are obtained
from a single \emph{base type} $\bool$ (for bool)
according to ${\sigma::=\bool\mid\sigma\sigma}$.  We
think of $\sigma\tau$ as the type of functions from
$\sigma$ to $\tau$.  \emph{Terms} ($s$, $t$, $u$) are
obtained from \emph{names} ($x$, $y$, $z$, $f$,~$g$)
according to $s::=x\mid\lambda x.s\mid s s$.  We assume
a \emph{typing relation $s:\sigma$} satisfying the
following properties:
\begin{enumerate}
\item For every term $s$ there is at most one type
  $\sigma$ such that $s:\sigma$.
\item For every type $\sigma$ there are infinitely many
  names $x$ such that $x:\sigma$.
\item For all $x$, $s$, $\sigma$, $\tau$: \
  $\lambda x.s:\sigma\tau\iff x:\sigma\land{s}:\tau$.
\item For all $s$, $t$, $\sigma$: \
  $s\,t:\sigma\iff\exists\tau\colon~s:\tau\sigma\land{t}:\tau$.
\end{enumerate}
A term $s$ is \emph{well-typed} if there is a type
$\sigma$ such that $s:\sigma$.  We only consider
well-typed terms.  We use \emph{$\Lambda$} to denote
the set of all well-typed terms.  We omit parentheses
according to
$\sigma\tau\rho\rightsquigarrow\sigma(\tau\rho)$ and
$stu\rightsquigarrow(st)u$.

Terms of type $\bool$ are called \emph{formulas}.  We
fix two names $\bot\colon\bool$ and
$\to\colon\bool\bool$ and call them \emph{constants}.
All other names are called \emph{variables}.  In a term
$\lam{x}s$, the bound name $x$ must be a variable.  We
use \emph{$\FV s$} to denote the set of all variables
that \emph{occur free in $s$}.  A term $s$ is
\emph{closed} if $\FV s=\emptyset$.

\emph{Contexts} are obtained according to
$C::=[]\mid\lambda x.C\mid{C}\,s\mid{s}\,C$.  The
notation $C[s]$ describes the term obtained by
replacing the hole~$[]$ of~$C$ with $s$ (capturing is
ok).  We assume a \emph{substitution operation $s^x_t$}
that yields for $s$, $x$,~$t$ a term that can be
obtained from $s$ by \emph{capture-free substitution}
of $t$ for~$x$, possibly after renaming of bound
variables.  \emph{Lambda equivalence $\siml$} is the
least equivalence relation on $\Lambda$ that satisfies
the following properties:
\begin{description}
\item[$\N{(\alpha)}$]~
  $\lam{x}{s}\siml\lam{y}{\Subst{s}{x}{y}}$~~ 
  if $y\notin\FV{s}$
\item[$\N{(\beta)}$]~
  $(\lam{x}s)t\siml\Subst{s}{x}{y}$~~
\item[$\N{(\eta)}$]~
  $\lam{x}{sx}\siml s$~~
  if $x\notin\FV{s}$
\item[$\N{(\gamma)}$]~
  if $s\siml{t}$, then $C[s]\siml C[t]$
\end{description}
It is easy to see that $\alpha$ is subsumed by the
other properties.  A term $s$ is a \emph{subterm} of a
term $t$ if there exists a context $C$ such that
$C[s]=t$.  A \emph{$\beta$-redex} is a term of the form
$(\lam{x}s)t$.  A term is \emph{$\beta$-normal} if none
of its subterms is a $\beta$-redex.  The following fact
is well-known \cite{HindleySeldin}.

\begin{prop}
  \label{prop:beta-nf}
  For every term $s$ there exists a lambda equivalent
  term $t$ such that $t$ is $\beta$-normal and
  satisfies $\FV{t}\subseteq\FV{s}$.
\end{prop}

An \emph{interpretation} is a function $\mci$ that maps
every type to a nonempty set and every name $x:\sigma$
to an element of $\mci\sigma$.  We require
$\mci\bool=\{0,1\}$, $\mci\bot=0$ and $\mci(\to)a
b=\Cond{a{=}0}{1}{b}$ for all $a,b\in\{0,1\}$.  We will
only consider standard interpretations, that is,
interpretations that map a functional type $\sigma\tau$
to the set of all total functions from $\mci\sigma$ to
$\mci\tau$.

Every interpretation $\mci$ can be extended uniquely to
a function $\hat\mci$ that maps every term $s:\sigma$
to an element of $\mci\sigma$ and treats applications
and abstractions as one would expect.  An
interpretation \emph{$\mci$ satisfies a formula $s$}
if $\hat\mci s=1$.  A formula is \emph{valid} if it is
satisfied by every interpretation.

\section{Denotational Completeness}

We fix an interpretation $\mcb$.  We have
$\mci\sigma=\mcb\sigma$ for every
interpretation $\mci$ and every type $\sigma$.
Moreover, we have $\hat\mci{s}=\hat\mcb{s}$ for every
closed term~$s$ and every  interpretation
$\mci$.  We will show that our language is
\emph{denotationally complete}, that is, for every type
$\sigma$ and every value $a\in\mcb\sigma$ there is a
closed term $s:\sigma$ such that $\hat\mcb{s}=a$.

It is well-known that implication and falsity can
express the usual propositional connectives:
\begin{align*}
  \N{\top}&~~:=~~\bot\to\bot & \N{s\lor t}&~~:=~~(s\to t)\to t\\
  \N{\neg s}&~~:=~~s\to\bot & \N{s\land t}&~~:=~~\neg(\neg s\lor\neg t)\\
  && \N{s\equiv t}&~~:=~~(s\to t)\land(t\to s)
\end{align*}
Note that Boolean equivalence $s\equiv{t}$ is Boolean
identity (i.e., $\mci{s}=\mci{t}$ iff $\mci$ satisfies
$s\equiv{t}$).  Notationally, we assume the operator
precedence ${\equiv},~{\to},~{\lor},~{\land},~\neg$
where $\neg$ binds strongest.

To show denotational completeness, we will define a
family of \emph{quote functions}
$\N{\Quote_\sigma}:\mcb\sigma\to\Lambda_0^\sigma$ where
\emph{$\Lambda_0^\sigma$} is the set of all closed
terms of type $\sigma$.  The quote functions will
satisfy $\hat\mcb(\Quote_\sigma{a})=a$ for all
$a\in\mcb\sigma$ and all types $\sigma$.  

The quote functions are defined by recursion on types.
The definition of the basic quote function
$\Quote_\bool$ is straightforward.  To explain the
definition of the other quote functions, we consider
the special case $\Quote_{\sigma\bool}$.  We start with
\begin{align*}
  \Quote_{\sigma\bool}(a)
  &~=~\lambda x.\bigvee_{
      \begin{subarray}{c}
        b\in\mcb\sigma \\[1pt]
        a b=\True
      \end{subarray}}
  x\mathrel{\deq}_\sigma(\Quote_\sigma{b})
\end{align*}
It remains to define a closed term $\deq_\sigma$ that
denotes the identity predicate for~$\mcb\sigma$.  If
$\sigma=\bool$, $\lambda x y.~x\equiv{y}$ does the job.
If $\sigma=\sigma_1\sigma_2$, we rely on recursion and
define
\begin{align*}
  \deq_\sigma
  &~=~\lambda fg.~\bigwedge_{\valv\in\mcb\sigma_1}
  f(\Quote_{\sigma_1}b)\mathrel{\deq}_{\sigma_2}g(\Quote_{\sigma_1}b)
\end{align*}
Figure~\ref{fig:quote} shows the full definition of the
quote functions. 
\begin{figure}
  \begin{align*}
    \Quote_{\sigma_1\ldots \sigma_n\bool}\,\valv~
    &:=~\lambda x_1\ldots x_n.\!\!
    \bigvee_{\begin{subarray}{c}
        \langle\valw_i\in\mcb \sigma_i\rangle\\[1pt]
        \valv \valw_1\ldots \valw_n=\True
      \end{subarray}}
      \bigwedge_{1\le j\le n}x_j\deq_{\sigma_j}(\Quote_{\sigma_j}\valw_j)
    & \N{\text{D}{\Quote}}\\[2mm]
    {\forall_\sigma}~
    &:=~\lambda f.\bigwedge_{\valv\in\mcb\sigma}f(\Quote_{\sigma}\valv)
    & \N{\text{D}\forall}\\[1mm]
    {\deq_\bool}~ &:=~\lambda x y.~x\equiv y 
    & \N{\text{D}{\deq}}\\[2mm]
    {\deq_{\sigma\tau}}~&:=~\lambda f g.~\forall_{\!\sigma}
    (\lambda x.~f x\deq_\tau g x)
    & \N{\text{D}{\deq}}
  \end{align*}
  \begin{center}\small
    $\langle\valw_i\in\mcb \sigma_i\rangle$ stands for
    $(\dd{b}n)\in\mcb\sigma_1\times\dots\times\mcb\sigma_n$
  \end{center}
  \medskip
  \caption{Quote Functions $\Quote_\sigma$ and Notations
    $\forall_\sigma$ and $\deq_\sigma$}
  \label{fig:quote}
  \bigskip
\end{figure}
A disjunction with an empty index set denotes $\bot$,
and a conjunction with an empty index set denotes
$\top$.  The notations \emph{$\deq_\sigma$} and
\emph{$\forall_\sigma$} defined in the figure will be
used in the following.  We write \emph{$\forall_\sigma
  x.\,s$} for $\forall_\sigma(\lambda x.s)$.  The
notational operator $\deq_\sigma$ will be used with a
precedence higher than $\neg$ (i.e,
$\neg\,s\,{\deq_\sigma}t$ stands for $\neg(s\deq_\sigma t)$).  The
following results are all straightforward consequences
of the definitions in Figure~\ref{fig:quote}.

\begin{prop}
  \label{prop:defs-closed}
  The terms ${\Quote_\sigma}\valv$, $\forall_\sigma$
  and $\deq_\sigma$ are closed.
\end{prop}

\begin{prop} \label{prop:soundness-defs}
  Let $\sigma$ be a type, $f\in\mcb(\sigma\bool)$, and
  $\valv,\valw\in\mcb\sigma$. Then:
  \begin{enumerate}
  \item $\hat\mcb({\Quote_\sigma}\valv)=\valv$ 
  \item $\hat\mcb({\forall_\sigma})f=\True\iff 
    \forall\valu\in\mcb\sigma\colon~f\valu=\True$ 
  \item $\hat\mcb({\deq_\sigma})\valv \valw=\True\iff\valv=\valw$
  \end{enumerate}
\end{prop}

Note that statement~(1) of
Proposition~\ref{prop:soundness-defs} implies that our
language is denotationally complete.  We state this
important fact explicitly.

\begin{prop}[\N{Denotational Completeness}]
  Let $\sigma$ be a type and ${\valv\in\mcb\sigma}$. Then
  there is a closed term $s$ such that $\hat\mcb
  s=\valv$.
\end{prop}

\section{Proof System}

A \emph{sequent} is a pair \emph{$A\seq{s}$} where $A$
is a finite set of formulas and $s$ is a formula.  The
letter~$A$ will always denote a finite set of formulas.
We write \emph{$\FV{A}$} for the set of all variables
that occur free in at least one formula in $A$.  An
interpretation \emph{$\mci$ satisfies $A$} if it
satisfies every formula in $A$.  A sequent $A\seq{s}$
is \emph{valid} if every interpretation that satisfies
$A$ also satisfies $s$.  A context $C$ \emph{captures}
a variable~$x$ if the hole of $C$ is in the scope of a
binder $\lambda x$.  A context~$C$ is \emph{admissible
  for $A$} if it does not capture any variable in $\FV
A$.

Figure~\ref{fig:proof-system} defines a sequent-based
proof system for our language.
\begin{figure}
  \begin{mathpar}
    \inferrule*[left=\emph{\textup{Triv}}~~]{~}{A,s\seq{s}}
    \and
    \inferrule*[left=\emph{\textup{Weak}}~~]{A\seq t}{A,s\seq t}
    \and
    \inferrule*[left=\emph{\textup{Ded}}~~]{A,\,s\seq t}{A\seq s\to t}
    \and
    \inferrule*[left=\emph{\textup{MP}}~~]{A\seq s\to t \\ A\seq s}{A\seq t}
    \and
    \inferrule*[left=\emph{\textup{DN}}~~]{A\seq\neg\neg s}{A\seq s}
    \\
    \inferrule*[left=\emph{\textup{Lam}}~~,right=~~$s\siml t$]
    {A\seq s}{A\seq t}
    \and
    \inferrule*[left=\emph{\textup{BR}}~~,right=~~C\textup{ admissible for }A]
    {A\seq s\equiv t \\ A\seq C[s]}{A\seq C[t]}
 \end{mathpar}
  \caption{Proof System}
  \label{fig:proof-system}
  \bigskip
\end{figure}
The first five rules (Triv, Weak, Ded, MP, DN) are
well-known from propositional logic.  We refer to the
proof system obtained with these rules as the
\emph{propositional subsystem}.  The propositional
subsystem differs from a pure propositional system in
that the propositional variables may be instantiated
with any term of type $\bool$.  The rule Lam
incorporates lambda equivalence.  The final rule BR
provides for replacement with respect to Boolean
equations.  A replacement may capture variables of the
equation if they don't occur in the assumptions.  A
sequent is \emph{deducible} if it is derivable with the
proof rules.  A formula $s$ is \emph{deducible} if the
sequent $\emptyset\seq{s}$ is deducible.

\begin{prop}[\N{Soundness}]
  Every deducible sequent is valid.
\end{prop}

\begin{pf}
  It suffices to show that every instance of every
  proof rule is sound, that is, that the conclusion is
  valid if all the premises are valid.  This is obvious
  for the propositional rules and well-known for Lam.
  The soundness of BR can be shown by induction on the
  context $C$.
\end{pf}

Let us look at an example illustrating that the
completeness of the proof system is not obvious.
Consider the formula $f(f(f x))\equiv f x$, where
$x:\bool$ and $f:\bool\bool$ are variables.  The
formula is valid.  Checking this claim is easy since
there are only 4 functions of type $\bool\bool$
(negation, the identity function, and the two constant
functions).  But for non-experts, a proof of the formula in
our proof system is not obvious.

A \emph{propositional formula} is a formula $s$ that
can be obtained according to
$s::=x\mid\bot\mid{s}\to{s}$ where $x$ serves as a
placeholder for variables.  A \emph{tautology} is a
valid propositional formula.  A formula is
\emph{tautologous} if it is a substitution instance of
a tautology.

\begin{prop}[\N{Taut}]
  Every tautologous formula is deducible.
\end{prop}

\begin{pf}
  We take it for granted that the propositional
  subsystem can deduce every tautology.  Since the
  instances of the propositional proof rules are closed
  under substitution of variables, propositional proof
  trees are closed under substitution of variables.
  Hence the claim follows. 
\end{pf}
  
We use \emph{$\vdash$} to denote the set of all
deducible sequents.  Since sequents are pairs,~$\vdash$
is a binary relation.  We 
write \emph{$A\vdash{s}$} if the sequent $A\seq{s}$ is
deducible, and \emph{$\vdash{s}$} if the formula $s$ is
deducible.  The next proposition states properties of
$\vdash$ that we will use in the following.

\begin{prop}
  ~
  \begin{description}
  \item[\N{Ded}]~ $A\vdash{s}\to{t}\iff A,s\vdash{t}$ 
  \item[\N{Cut}]~ $\displaystyle
    A,s_1,\ldots,s_n\vdash{s}~\land~
    A\vdash{s_1}~\land~\dots~\land~A\vdash{s_n}
    ~\impl~
    A\vdash{s}$ 
  \item[\N{And}]~ $s_1,s_2\vdash{s_1}\land{s_2}$ 
  \item[\N{Ref}]~ $A\vdash{s}\equiv{s}$ 
  \item[\N{Sub}]~ $A\vdash{s}~\impl~\Subst{A}xt\vdash\Subst{s}xt$
  \end{description}
\end{prop}

\begin{pf}
  The derivation of Ded and Cut is straightforward.
  And follows with Taut and Ded since
  $x\to{y}\to{x}\land{y}$ is a tautology.  Ref follows
  with Taut since $x\equiv{x}$ is a tautology.  Because
  of Ded it suffices to show Sub for $A=\emptyset$.
  Let $\vdash{s}$.  Then $\vdash{s}\equiv\top$ by Taut
  and BR since $x\equiv(x\equiv\top)$ is a tautology.
  By Ref an BR we obtain
  $\vdash(\lam{x}s)t\equiv(\lam{x}\top)t$.  Thus
  $\vdash\Subst{s}xt\equiv\top$ by Lam.  Hence
  $\vdash\Subst{s}xt$ by Taut and~BR. 
\end{pf}

\begin{prop} 
  \label{lem:closed-fml-prop}
  Every closed and $\beta$-normal formula is
  propositional.
\end{prop}

\begin{pf}
  By induction on the size of formulas.  Let $s$ be a
  $\beta$-normal and closed formula. Then $s=x s_1\dots
  s_n$ where $\dd{s}{n}$ are all closed and
  $\beta$-normal. Since $s$ is closed, either $x=\bot$
  or $x={\to}$.  If $x=\bot$, then $n=0$ and hence $s$
  is propositional.  If $x={\to}$, then $n=2$ and the
  claim follows by the induction hypothesis applied to
  $s_1$ and $s_2$. 
\end{pf}

\begin{prop}[\N{Closed Completeness}]
  \label{prop:closed-comp}~\\
  Every closed and valid formula is deducible.
\end{prop}

\begin{pf}
  By Proposition~\ref{prop:beta-nf}, Lam, and Soundness
  it suffices to show the claim for closed, valid,
  $\beta$-normal formulas.  By
  Proposition~\ref{lem:closed-fml-prop} we know that
  such formulas are tautologies.  Now the claim follows
  with Taut. 
\end{pf}

\section{Deductive Completeness} 
\label{sec:compl-high-types}

We say that $\vdash$ is \emph{complete} if every valid
formula is deducible.  By Ded, completeness of $\vdash$
implies that every valid sequent is deducible.  For
every type $\sigma$, we define three properties:
\begin{description}
\item[\N{All$_\sigma$}] For all $f:\sigma\bool$ and
  $x:\sigma$ it holds:~ $\forall_\sigma{f}\vdash{fx}$
\item[\N{Enum$_\sigma$}] For all $x:\sigma$ it holds:~
  $\displaystyle\vdash\bigvee_{a\in\mcb\sigma}(\Quote_\sigma
  a)\deq_\sigma x$
\item[\N{Rep$_\sigma$}] For all $A$ and all formulas
  $s\deq_\sigma t$ and $C[s]$ such that the context $C$
  is admissible for $A$, it holds:~ If $A\vdash
  s\deq_\sigma t$ and $A\vdash C[s]$, then $A\vdash
  C[t]$.
\end{description}
We will show that the properties hold for all types.

\begin{lem} 
  \label{lem:comp-crit}
  If All$_\sigma$ holds for all types $\sigma$, then\/ $\vdash$ is
  complete.
\end{lem}

\begin{pf}
  Assume All$_\sigma$ holds for all types $\sigma$.  Let $s$ be
  a valid formula.  We show $\vdash s$.  Let $\FV
  s=\{\dd{x}{n}\}$.  Then $\forall x_1\dots\forall
  x_n.s$ is closed and valid.  Hence, ${\vdash\forall
    x_1\dots\forall x_n.s}$ by
  Proposition~\ref{prop:closed-comp}.  The claim now
  follows by repeated application of All$_\sigma$, Sub,
  Lam, and Cut. 
\end{pf}

\begin{lem} 
  \label{lem:prop-rei}
  $\vdash\Quote_\tau(ab)\deq_\sigma(\Quote_{\sigma\tau}a)(\Quote_{\sigma}b)$
\end{lem}

\begin{pf}
  Since the formula is closed, by
  Proposition~\ref{prop:closed-comp} it suffices to
  show that it is valid.  This holds since
  $\hat\mcb(\Quote_\tau(ab))= ab=
  (\hat\mcb(\Quote_{\sigma\tau}a))(\hat\mcb(\Quote_{\sigma}b))=
  \hat\mcb((\Quote_{\sigma\tau}a)(\Quote_{\sigma}b))$
  by Proposition~\ref{prop:soundness-defs}. 
\end{pf}

\begin{lem}
  \label{lem:distrib}
  Let $I$ and $J$ be finite sets and $x_{i,j}:\bool$ be
  a variable for all\/ $i\in{I}$, $j\in{J}$.  Moreover,
  let\/ $\fto{I}J$ be the set of all total functions
  $I\to{J}$.  Then:
  \begin{equation*}
    \displaystyle\vdash\bigwedge_{i\in I}\,\,\bigvee_{j\in J}\,x_{i,j}
    ~\equiv\bigvee_{\varphi\in\fto{I}{J}}\,\,\bigwedge_{i\in I}\,x_{i,\varphi i}
  \end{equation*}
\end{lem}

\begin{pf}
  Let $s$ and $t$ be the left and the right term of the
  equivalence in question, respectively, and let $\mci$ be an
  interpretation. The claim follows by Taut if we can show that
  $\hat\mci{s}=1$ iff $\hat\mci{t}=1$.  Let $\hat\mci{s}=1$.  Then for
  every $i\in{I}$ there exists a $j\in{J}$ such that
  $\mci(x_{i,j})=1$.  Hence there exists a function
  $\varphi\in\fto{I}{J}$ such that
  $\mci(x_{i,\varphi{i}})=1$ for every $i\in{I}$.
  Hence $\hat\mci{t}=1$.  The other direction follows
  analogously. 
\end{pf}

\begin{lem} 
  \label{lem:crit-crit}
  Let\/ $\sigma$ be a type. If Enum$_\sigma$ and Rep$_\sigma$ hold, then
  All$_\sigma$ holds.
\end{lem}

\begin{pf}
  Assume that Enum$_\sigma$ and Rep$_\sigma$ hold.  Let
  $f:\sigma\bool$, $x:\sigma$, and $a\in\mcb\sigma$. By
  Taut we have:
  \[\vdash\left(\bigwedge_{b\in\mcb\sigma}f(\Quote_\sigma b)\right)\to
  f(\Quote_\sigma a)\]
  Hence, by D$\forall$, Lam, and Weak:
  \[(\Quote_\sigma a)\deq_\sigma x\vdash\forall_\sigma f\to f(\Quote_\sigma a)\]
  Hence, by Triv, Rep$_\sigma$, and Ded:
  \[\vdash(\Quote_\sigma a){\deq_\sigma}x\to\forall_\sigma f\to f x\]
  Since $a\in\mcb\sigma$ was chosen freely, we have by
  And and Cut:
  \[\vdash\bigwedge_{a\in\mcb\sigma}
  ((\Quote_\sigma a){\deq_\sigma}x\to\forall_\sigma f\to f x)\]
  Now, by Taut and MP, it follows:
  \[\vdash\left(\bigvee_{a\in\mcb\sigma}(\Quote_\sigma a)\deq_\sigma x\right)\to
  \forall_\sigma f\to f x\]
  The claim follows by Enum$_\sigma$ and MP. 
\end{pf}

\begin{lem} 
  \label{lem:compl}
  Enum$_\sigma$ and Rep$_\sigma$ hold for all types
  $\sigma$.
\end{lem}

\begin{pf}
  By induction on $\sigma$.
  
  We first show Enum$_\sigma$.  Let $x:\sigma$.  If
  $\sigma=\bool$, the claim follows by Taut.
  Otherwise, let $\sigma=\sigma_1\sigma_2$.  By
  Enum$_{\sigma_2}$ (induction hypothesis) and Sub, we
  obtain
  \[
  \vdash\bigvee_{c\in\mcb\sigma_2}
  (\Quote_{\sigma_2}c)\deq_{\sigma_2}x(\Quote_{\sigma_1} b)
  \] 
  for every $b\in\mcb\sigma_1$.  Hence, by And:
  \[\vdash\bigwedge_{b\in\mcb{\sigma_1}}\bigvee_{c\in\mcb\sigma_2}
  (\Quote_{\sigma_2}c)\deq_{\sigma_2}x(\Quote_{\sigma_1} b)\]
  By Lemma~\ref{lem:distrib}, Sub, and BR, this yields:
  \[
  \vdash\bigvee_{a\in\mcb\sigma}\bigwedge_{b\in\mcb\sigma_1}
  (\Quote_{\sigma_2}(ab))\deq_{\sigma_2}x(\Quote_{\sigma_1} b)
  \]
  By repeated application of Lemma~\ref{lem:prop-rei} and
  Rep$_{\sigma_2}$ (induction hypothesis), we
  obtain
  \[
  \vdash\bigvee_{a\in\mcb\sigma}\bigwedge_{b\in\mcb\sigma_1}
  (\Quote_{\sigma} a)(\Quote_{\sigma_1} b)\deq_{\sigma_2}x(\Quote_{\sigma_1} b)
  \]
  which is Enum$_\sigma$ up to D$\deq$, D$\forall$ and
  Lam.

  Next we show Rep$_\sigma$.  Let $A\vdash s\deq_\sigma
  t$ and $A\vdash C[s]$, and let $C$ be admissible for
  $A$.  We show $A\vdash C[t]$.  If $\sigma=\bool$, the
  claim is immediate by BR.  Otherwise, let
  $\sigma=\sigma_1\sigma_2$.  By Triv, D$\deq$, and Lam
  we have:
  \[
  s\deq_\sigma t\vdash\forall_{\sigma_1}y.~s y\deq_{\sigma_2}t y
  \]
  for some variable $y\notin\FV(s{\deq_\sigma}t)$.  By
  the induction hypothesis and
  Lemma~\ref{lem:crit-crit}, we have All$_{\sigma_1}$.
  By Sub, Lam, Weak, and Cut we then obtain:
  \[s\deq_\sigma t\vdash s y\deq_{\sigma_2}t y\]
  Hence, by Weak, the assumption $A\vdash s\deq_\sigma t$, and Cut:
  \[
  A\vdash s y\deq_{\sigma_2}t y
  \] 
  Since $A\vdash C[s]$ and $y\notin\FV{s}$, we have by
  Lam:
  \[
  A\vdash C[\lambda y.s y]
  \] 
  Since the context $C[\lam{y}{[]}]$ is admissible for
  $A$ and Rep$_{\sigma_2}$ holds by the induction
  hypothesis, we have:
  \[A\vdash C[\lambda y.t y]\]
  The claim follows by Lam. 
\end{pf}

\begin{thm}[Deductive Completeness]
  $\vdash$ is complete.
\end{thm}

\begin{pf}
  Follows by
  Lemmas~\ref{lem:compl},~\ref{lem:crit-crit},
  and~\ref{lem:comp-crit}. 
\end{pf}

\bibliography{paper}
\bibliographystyle{acm}

\end{document}